\documentclass[pra,nofootinbib]{revtex4}
\usepackage{amssymb,amsmath}
\usepackage{epsfig}
\def\shiftleft#1{#1\llap{#1\hskip 0.04em}}
 
\def\shiftdown#1{#1\llap{\lower.04ex\hbox{#1}}}
\def\thick#1{\shiftdown{\shiftleft{#1}}}
\def\b#1{\thick{\hbox{$#1$}}}
\begin{document}
\title{Scattering processes in antiprotonic hydrogen - hydrogen atom collisions}
\thanks{The work was supported by Russian Foundation for Basic Research, grant
No 03-02-16616.}
\author{V.P. Popov and V.N. Pomerantsev }
\affiliation{Institute of Nuclear Physics, Moscow State
University}
%\date{}

\begin{abstract}{The elastic scattering, Stark transitions and
Coulomb deexcitation of excited antiprotonic
hydrogen atom in collisions with hydrogenic atom have been
studied in the framework of the fully quantum-mechanical
close-coupling method for the first time.  The total cross
sections $\sigma_{nl \rightarrow n'l'}(E)$ and averaged on
the initial angular momentum $l$ cross sections
$\sigma_{n\rightarrow n'}(E)$ have been calculated for the
initial states of $(\bar{p}p)_{n}$ atoms with the principal
quantum number $n=3 - 14 $ and at the relative energies
$E=0.05 - 50$~eV. The energy shifts of the $ns$ states due
to the strong interaction and relativistic effects are taken
into account. Some of our results are compared with the semiclassical 
calculations.}
\end{abstract}

\maketitle

\section{Introduction}

The slowing down and Coulomb capture of the negative
particle $M^{-}$ ($M^{-}=\mu^{-},\pi^{-}, etc.$) in
hydrogen media lead to the formation of the
$M^{-}$-molecular complex the decay of which  results in the exotic
atom formation in highly excited states with the principal
quantum number $n \sim \sqrt{\mu}$ where $\mu$ is the
reduced mass of the exotic atom. Their initial
$nl$-population and kinetic energy distribution of the exotic atom are defined
by the competition of different decay modes of this
complex. The further destiny of the exotic atom depends on
the kinetics of the processes occurring in the deexcitation
cascade. The experimental data are mainly appropriate to
the last stage of the atomic cascade, such as X-ray yields
and the products of the weak or strong interaction of the
exotic particle in the low angular momentum states with
hydrogen isotopes.

 Hadronic hydrogen atoms are of special interest among exotic
atoms because they have the simplest structure and are the
probe in the investigations of the various aspects of both
the exotic atom physics and the elementary hadron-nucleon interactions at zero energy. 
In particular, in order to analyze the precision spectroscopy experimental
data~\cite{1} the kinetic energy distribution must be taken 
into account. The velocity of the exotic hydrogen atom plays an important
role due to the effect of the Stark transitions on the L X-rays
yields and the Doppler broadening of the L-lines owing to the
preceding Coulomb deexcitation transitions. The energy
release in the last process leads to an acceleration of 
colliding partners. So the reliable theoretical backgrounds
on the processes both in low-lying and in highly excited states are required for
the detailed and proper analysis of these data.

In this paper we present the first step to {\em ab initio}
quantum-mechanical treatment of non-reactive scattering
processes of the excited antiprotonic hydrogen atom in collisions with the
hydrogenic atom in the ground state:
\begin{itemize}
\item[-] elastic scattering
\begin{equation}
(ax)_{n l} + (be^-)_{1s} \to (ax)_{n l} + (be^-)_{1s};
\end{equation}
\item[-] Stark mixing
\begin{equation}
(ax)_{nl} + (be^-)_{1s} \to (ax)_{nl'} + (be^-)_{1s};
\end{equation}
\item[-] Coulomb deexcitation
\begin{equation}
(ax)_{nl} + (be^-)_{1s} \to (ax)_{n'l'} + (b e^-)_{1s}.
\end{equation}
\end{itemize}
Here $(a,b) = (p,d,t)$ are hydrogen isotopes and $x=K^-,
\tilde{p}$; $(n,l)$ and $1s$ are the principal and orbital quantum
numbers of the exotic and hydrogenic atom, respectively.
 The processes (1) - (2) decelerate and
accelerate while the Coulomb deexcitation (3) accelerates the
exotic atoms, influencing their quantum number and energy
distributions during the cascade. The last process has
attracted  particular attention especially after the "hot"
$\pi p$ atoms with the kinetic energy up to 200 eV were
found experimentally\cite{2}. Due to the similarity of the general features of
the exotic atoms the Coulomb deexcitation process must be also taken into
account for the other exotic atoms.

 Starting from the classical paper by Leon and Bethe~[3],
 Stark transitions have been treated in the semiclassical
 straight-line trajectory approximation (see [4] and
 references therein). The first fully quantum-mechanical treatment
 of the processes (1) - (2) based on the adiabatic description
 was given in [5-8]. Recently\cite{9, 10} the elastic scattering
 and Stark transitions (for $n=2-5$) have also been studied
in a close-coupling approach treating the interaction of
the exotic hydrogen atom with the hydrogenic one in the dipole
approximation with electron screening taken into account by the 
model. As for higher exotic atom states $(n>5)$, the
semiclassical approach~\cite{10}  is
used for the description of these processes.

 As concerning the acceleration process (3) in the muonic and hadronic
 hydrogen atoms, the parameterization based on the
 calculations in the semiclassical model~\cite{11} (see also [13])
 is used for the low-lying states ($n=3-7$) and the
 results of the classical-trajectory Monte Carlo
 model~\cite{12, 13} are used for higher exotic atom states 
 in the cascade calculations.

The main aim of this paper is to obtain the cross sections
of the processes (1)-(3) for the excited antiprotonic atom
beginning from the low collision energies in the framework of the fully
quantum-mechanical approach. For this purpose the
unified treatment of the elastic scattering, Stark transitions
and Coulomb deexcitation within the close-coupling method has been used.
This approach has been recently applied for the study of the
differential and total cross sections of the elastic
scattering, Stark transitions and Coulomb deexcitation in
the collisions of the excited muonic~\cite{14} and pionic~\cite{15} 
hydrogen atoms with the hydrogen ones. In the following Section 
we briefly describe the close-coupling formalism.
The results of the close-coupling  calculations concerning
the total cross sections of the processes (1)-(3) are presented
and discussed in Section III. Finally, summary and concluding
remarks are given in Section IV.

\section{Close-coupling approach}

\subsection{ Total wave function of the binary system
in terms of basis states}

The total wave function $\Psi (\boldsymbol{\rho},
\mathbf{r}, \mathbf{R})$ of the four-body system $(a
\bar{p} + b e^{-})$  satisfies the time independent
Schr\"{o}dinger equation with the Hamiltonian which after
separating the center of mass motion can be written as
\begin{equation}
H = -\frac{1}{2m}\Delta _{\mathbf{R}} +
h_{\bar{p}}(\boldsymbol{\rho}) + h_e(\mathbf{r})
+V(\mathbf{r},\boldsymbol{\rho},\mathbf{R})
 \label{Ham}
\end{equation}
($m$ is the reduced mass of the system). Here we use the 
set of Jacobi coordinates $(\mathbf{R},
\boldsymbol{\rho},  \mathbf{r})$:
\[\mathbf{R} =
\mathbf{R}_H - \mathbf{R}_{\bar{p}a},\quad
\boldsymbol{\rho} = \mathbf{r}_{\bar{p}} -
\mathbf{r}_a,\quad \mathbf{r}= \mathbf{r}_e
-\mathbf{r}_b,\] where $\mathbf{r}_a, \mathbf{r}_b,
\mathbf{r}_{\bar{p}}, \mathbf{r}_e$ are the radius-vectors
of the nuclei, muon and electron in the lab-system and
$\mathbf{R}_H, \mathbf{R}_{\bar{p} a}$ are the center of
mass radius-vectors of hydrogen and exotic atoms,
respectively. The Hamiltonians $h_{\bar{p}}$ and $h_e $ of
the free exotic and hydrogen atoms, respectively, satisfy
the Schr\"{o}dinger equations
\begin{align}
h_\mu \Phi_{nlm}(\boldsymbol{\rho}) &= \varepsilon_{nl}
\Phi_{nlm}(\boldsymbol{\rho}),
\\
h_e\varphi_{n_e l_e}(\mathbf{r}) &=
\epsilon_{n_e}\varphi_{n_e l_e}(\mathbf{r}),
\end{align}
where $\Phi_{nlm}(\boldsymbol{\rho})$ and
$\varphi_{n_e}(\mathbf{r})$ are the hydrogen-like wave
functions of the exotic atom and hydrogen atom bound
states, $\varepsilon_{nl}$ and $\epsilon_{n_e}$ are the
corresponding eigenvalues. In the present study 
$\varepsilon_{nl}$ includes beyond the standard
non-relativistic two-body Coulomb problem the energy shifts
due to the strong interaction, vacuum polarization and finite
size. It is worthwhile noting that in order to treat the hadronic $ns$
states as normal asymptotic states in the scattering
problem we take into consideration only the real part of
the complex strong interaction energy shift.

The interaction potential 
\begin{equation}
V(\mathbf{r},\boldsymbol{\rho},\mathbf{R})=V_{ab}+V_{\mu
b}+V_{ae}+V_{\mu e} \label{}
\end{equation}
includes the two-body Coulomb interactions between the
particles from two colliding subsystems:
\begin{eqnarray}
 V_{ab}=\frac{1}{r_{ab}}=
 |\mathbf{R}+\nu \boldsymbol{\rho}-\nu_e \mathbf{r}|^{-1},
 V_{\bar{p} b}=
 -\frac{1}{r_{\bar{p} b}}=
 -|\mathbf{R}-\xi\boldsymbol{\rho}-\nu_e \mathbf{r}|^{-1},
 \\
 V_{\bar{p} e}=
 \frac{1}{r_{\bar{p} e}}=
 |\mathbf{R}-\xi \boldsymbol{\rho}+\xi_e \mathbf{r}|^{-1},
V_{ae}=-\frac{1}{r_{ae}}=-|\mathbf{R}+\nu
\boldsymbol{\rho}+\xi_e \mathbf{r}|^{-1},
\end{eqnarray}
where the following notations are used:
\begin{eqnarray}
\nu = m_{\bar{p}}/(m_{\bar{p}} +m_a),\; \xi = m_a
/(m_{\bar{p}}+m_a),
\\ \nu_e = m_e /(m_e+m_b), \; \xi_e = m_b /(m_e +m_b),
\end{eqnarray}
($m_a, m_b, m_{\bar{p}}$ and $m_e$ are the masses of 
hydrogen isotopes, antiproton and electron, respectively).
 Atomic units (a.u)
 $\hbar=e=m_e m_b/(m_e+m_b)=1$ will be used throughout the
 paper unless otherwise stated.\\

  In this paper, as well as in the previous studies [11, 12-15],
   we assume that the state of the target
 electron is fixed during the collision. The electron
 excitations can be taken into account in a straightforward
 manner.
In a space-fixed coordinate frame we built the basis states
from the eigenvectors of the operators  $ h_e, h_{\bar{p}},
\mathbf{l}^2, \mathbf{L}^2, \mathbf{J}^2, J_z$ and the
total parity $\pi$ with the corresponding eigenvalues
$\varepsilon_{1s}, \varepsilon_n, l(l+1), L(L+1), J(J+1),
M$ and $(-1)^{l+L}$, respectively:

\begin{equation}
|1s,n l, L:JM\rangle\equiv \frac{1}{\sqrt{4\pi}} R_{1s}(r)
R_{nl}(\rho) {\cal Y}_{lL}^{JM}(\hat {\b\rho}, \hat{\bf
R}), \label{}
\end{equation}
where
\begin{equation}
{\cal Y}_{l L}^{JM} (\hat {\b\rho}, \hat{\bf R})\equiv
\sum_{m \lambda}\langle l m L \lambda |JM \rangle
Y_{lm}(\hat{\b\rho}) Y_{L \lambda}(\mathbf{\hat R}).
 \label{}
\end{equation}
Here the orbital angular momentum $\bf l$ of $(a\mu)_{nl}$
is coupled with the orbital momentum $\bf L$ of the
relative motion to give the total angular momentum, $\bf
{J=l + L}$. The explicit form of the radial hydrogen-like
wave functions $R_{nl}(\rho)$ will be given below.

Then, for the fixed values of $J, M, \pi = (-1)^{l+L}$ the
exact solution of the Schr\"{o}dinger equation
\begin{equation}
(E - H) \Psi_{E}^{JM\pi}({\bf r}, {\b\rho}, {\bf R}) = 0,
\label{}
\end{equation}
is expanded as follows
\begin{equation}
\Psi_{E}^{J M \pi}(\mathbf{r}, {\b\rho}, \mathbf{R}) =
\frac{1}{R} \sum_{nl L}G_{nlL}^{J \pi}(R)|1s,n
l,L:JM\rangle, \label{}
\end{equation}
where the $G_{nlL}^{J \pi}(R)$ are the radial functions of
the relative motion  and the sum is restricted by the  $(l,
L)$ values to satisfy the total parity conservation.  This
expansion leads to the coupled radial scattering equations
\begin{equation}
\left(\frac{d^2}{dR^2} + k^2_{n} -
\frac{L(L+1)}{R^2}\right)G^{J \pi}_{nlL}(R) = 2m
\sum_{n'l'L'}W^{J}_{n'l'L', nlL}(R)\,G^{J \pi
}_{n'l'L'}(R), \label{cce}
\end{equation}
where
$k^{2}_{nl}=2m(E_{cm}+\varepsilon_{n_il_i}-\varepsilon_{nl})$
specifies the channel wave numbers; $E_{cm}$ and
$\varepsilon_{n_il_i}$ are relative motion energy and exotic atom
bound energy in the entrance channel, respectively.

The radial functions $G_{E, nlL}^{J \pi}(R)$ satisfy the
usual plane-wave boundary conditions at $R\rightarrow 0$
\begin{equation}
G_{E, n'l'L'}^{J \pi}(0)=0 (\sim R^{L+1}) \label{}
\end{equation}
and at asymptotic distances ($R\rightarrow \infty$)
\begin{equation}
G_{E, n'l'L'}^{J \pi}(R)\Rightarrow \frac{1}{\sqrt
{k_f}}\{\delta_{if} \delta_{nn'} \delta_{ll'}
\delta_{LL'}e^{-i(k_{i}R-L\pi/2)}- S^J(nl, L\rightarrow
n'l', L')e^{i(k_{f}R-L'\pi/2)}\}, \label{}
\end{equation}
where $k_i$, $k_f$ are the wave numbers of initial and
final channels and $S^J(nl, L\rightarrow n'l', L')$ is the
scattering matrix in the total angular momentum
representation. Here and below the indices of the entrance
channel and target electron state are omitted for brevity.

\subsection{Potential matrix}

Here we present the derivation of the exact matrix of the
interaction potentials involved in the close-coupling
calculations. The interaction potential matrix
$W^{J}_{n'l'L', nlL}$ coupling the asymptotic initial $(n l
L; J)$ and final $(n' l' L'; J)$  channels is defined by

\begin{align}
W^{J}_{n'l'L,nlL}(R)&=\frac{1}{4\pi}\int{\rm d} {\bf
r}\,{\rm d}{\b \rho} \,{\rm d} \hat{\bf R} R^2_{1s}
(r)R_{nl} (\rho)R_{n'l'}(\rho) \nonumber\\ &\times {\cal
Y}^{JM}_{lL} (\hat{\b \rho},\hat{\bf R})\, V({\bf r},{\b
\rho},{\bf R})\, '({\cal Y}^{JM}_{l'L'})^{^*} (\hat{\b
\rho},\hat{\bf R}),
\end{align}
 where the radial hydrogen-like wave functions
 are given explicitly  by
\begin{equation}
R_{nl}(\rho)=N_{nl}\left(\frac{2\rho}{n
a}\right)^l\exp(-\rho/na)
 \sum_{q=0}^{n-l-1}S_{q}(n,
 l)\left(\frac{2\rho}{n a}\right)^q \label{}
\end{equation}
($a$ is the Bohr' radius of the exotic atom in a.u.) with
\begin{equation}
N_{nl}=\left(\frac{2}{n a}\right)^{3/2}
\left[\frac{(n+l)!(n-l-1)!}{2n}\right]^{1/2}, \label{}
\end{equation}
and
\begin{equation}
S_{q}(n, l) = (-)^q \frac{1}{q!(n-l-1-q)!(2l+1+q)!}.
\label{}
\end{equation}
 Averaging $V({\bf r},{\b \rho},{\bf R})$ over $1s$-state of hydrogen atom
 leads to
\begin{align}
V({\bf R},{\b \rho})&=\frac{1}{4\pi}\int_{0}^{\infty}{\rm
d} {\bf r} R^2_{1s} (r)V({\bf r},{\b \rho},{\bf R})=
\nonumber\\ &= \frac{1}{\xi_e}\{U_{\nu,\xi_e}({\bf
 R},{\b \rho})-U_{-\xi,\xi_e}({\bf R},{\b \rho})\} -
\frac{1}{\nu_e}\{U_{\nu,\nu_e}({\bf R},{\b
\rho})-U_{-\xi,\xi_e}({\bf R},{\b \rho})\}.
\end{align}
Then we use the transformation
\begin{multline}
U_{\alpha,\beta}({\bf R}, {\b \rho})=(1+\frac{\beta}{|{\bf
R}+ \alpha\boldsymbol{\rho}|}) {\rm e}^{-\frac{2|{\bf
R}+\alpha\boldsymbol{\rho}|}{\beta}}\equiv
 \lim_{x\to 1}\left(1-\frac{1}{2}\frac{\partial}{\partial x}\right)
\beta\frac{{\rm e}^{-\frac{2x|{\bf
R}+\alpha\boldsymbol{\rho}|}{\beta}}} {|{\bf
R}+\alpha\boldsymbol{\rho}|}.
\end{multline}
which allows us to apply the additional theorem for the
spherical Bessel functions~\cite{16}
\begin{multline}
\frac{{\rm e}^{-\lambda |{\bf R}_1+{\bf r}_1|}}{|{\bf
R}_1+{\bf r}_1|}=
\frac{4\pi}{\sqrt{R_1r_1}}\sum_{t\tau}(-1)^t
Y^*_{t\tau}(\hat{\bf R}_1) Y_{t\tau}(\hat{\bf r}_1)\times
\\ \times \left\{ K_{t+1/2}(\lambda R_1)\,I_{t+1/2}(\lambda
r_1)\left |_{r_1 <R_1} + I_{t+1/2}(\lambda
R_1)\,K_{t+1/2}(\lambda r_1)\right |_{r_1 >R_1} \right \}
\end{multline}
($I_p(x)$ and $K_p(x)$ are the modified  spherical Bessel
functions of the first and third kind). Furthermore, by
substituting the Eqs.(20)-(25) into  Eq.(19) we can integrate
over the angular variables $(\bf {R}, \boldsymbol{\rho})$.
Finally,  applying
 the angular momentum algebra and integrating over $\rho$, we obtain:
\begin{align}
W^{J}_{nlL,n'l'L'}(R)&=(-1)^{J+l+l'}i^{l'+L'-l-L}\sqrt{\hat{l}\hat{l'}\hat{L}\hat{L'}}
\sum_{t=0}^{t_m}(l0l'0|t0)(L0L'0|t0)
\left\{\begin{array}{lll}
l&l'&t\\L'&L&J\end{array}\right\}\times \nonumber \\
&\times \left\{\frac{1}{\xi_e}\left[ (-1)^t W_t(R,\nu
,\xi_e;nl,n'l') - W_t (R,\xi,\xi_e;nl,n'l')\right] \right
.- \nonumber \\ & - \left .  \frac{1}{\nu_e}\left[ (-1)^t
W_t(R,\nu,\nu_e;nl,n'l') - W_t(R,\xi,\nu_e;nl,n'l')\right]
\right\}
\end{align}
($t_m$ is the maximum value of the allowed multipoles).
Here the next notations are used:
\begin{align}
W_t(R,\alpha,\beta;n l,n'l')&={\cal N}_{nl,n'l'}
\sum_{m_1=0}^{n-l-1}S_{m_1}(n,l)\left(\frac{2n'}{n+n'}\right)^{m_1}
\sum_{m_2=0}^{n'-l'-1} S_{m_2}(n',
l')\left(\frac{2n}{n+n'}\right)^{m_2} \times \nonumber \\
\times
&\left\{H_t(x)J_1^{t,s}(x,\lambda(n,n',\alpha,\beta))-
h_t(x)J_2^{t,s}(x,\lambda(n,n',\alpha,\beta)) + \right .
\nonumber \\
&+F_t(x)J_3^{t,s}(x,\lambda(n,n',\alpha,\beta)) \left .
+f_t(x)J_4^{t,s}(x,\lambda(n,n',\alpha,\beta)) \right\},
\end{align}
where $x=2R/\beta $, $ s=l+l'+m_1+m_2 $, $ \hat{L}\equiv
2L+1$;
\begin{equation}
{\cal N}_{nl,n'l'} =
\frac{1}{n+n'}\!\left(\frac{2n'}{n+n'}\right)^{l+1}\!\left(\frac{2n}{n+n'}\right)^{l'+1}
\!\sqrt{(n+l)!(n-l-1)!(n'+l')!(n'-l'-1)}; \label{}
\end{equation}
\begin{equation}
\lambda(n,n',\alpha,\beta)=\frac{2nn'}{n+n'}\frac{a\alpha}{\beta};
\label{}
\end{equation}
\begin{equation}
H_t(x)=(1-2t)h_t(x)+xh_{t+1}(x); \label{}
\end{equation}
\begin{equation} F_t(x)=(1-2t)f_t(x)-xf_{t+1}(x). \label{}
\end{equation}
The functions $h_t(x)$ and $f_t(x)$ are given by
\begin{equation}
h_t(x)\equiv\sqrt{\frac{2}{\pi x}}K_{t+1/2}(x) \label{}
\end{equation}
and
\begin{equation}
f_t(x)\equiv\sqrt{\frac{\pi}{2 x}}I_{t+1/2}(x). \label{}
\end{equation}

The radial integrals $J_i^{t,s}(x,\lambda)$ are defined as
follows:
\begin{equation}
J_1^{t,s}(x,\lambda)=\int_{0}^{x/\lambda}
y^{s+2}e^{-y}f_t(\lambda y)\,{\rm d}y,
\end{equation}
\begin{equation}
J_2^{t,s}(x,\lambda)=\lambda J_1^{t+1,s+1}(x,\lambda),
\end{equation}
\begin{equation}
J_3^{t,s}(x,\lambda)=\int_{x/\lambda}^{\infty}
y^{s+2}e^{-y}h_t(\lambda y)\,{\rm d}y,
\end{equation}
\begin{equation}
J_4^{t,s}(x,\lambda)=\lambda J_3^{t+1,s+1}(x,\lambda)
\end{equation}
and calculated analytically using  the power series
 for the modified Bessel functions.

\subsection {Cross sections}

The transition amplitude from the initial state
$|nlm>$ to the final state $|n'l'm'>$ of the exotic atom can be
defined by
\begin{align}
f(nlm\rightarrow n'l'm'|\mathbf{k}_i, \mathbf{k}_f
\rangle&= \frac{2\pi
i}{\sqrt{k_{i}k_{f}}}\sum_{JMLL'\lambda
\lambda'}i^{L'-L}\langle lmL\lambda |J M\rangle \langle
l'm'L'\lambda'|J M\rangle \times \nonumber \\ &\times
Y_{L\lambda}^{*}(\mathbf{\hat k_i})
Y_{L'\lambda'}(\mathbf{\hat k_f})T^J(nlL\rightarrow
n'l'L'). \label{}
\end{align}
Here, $k_i$ and $k_f$ are the center of mass relative momenta in the
initial and final channels; $\mathbf{\hat k_i}$ and
$\mathbf{\hat k_i}$ are their unit vectors in the
space-fixed system, respectively, and, finally, the
transition matrix $T^J(nlL\rightarrow n'l'L')$ used here is
given by
\begin{equation}
T^J(nl, L\rightarrow n'l',
L')=\delta_{nn'}\delta_{ll'}\delta_{LL'}\delta_{mm'}\delta_{\lambda\lambda'}
- S^J(nl, L\rightarrow n'l', L'). \label{}
\end{equation}

In terms of the scattering amplitude (38) defined above, all the
types of both the differential and total cross sections of the all 
processes under consideration  for the transition from the
initial  $(n l)$ state to the final $(n'l')$ state  are
defined as:
\\ differential cross sections
\begin{equation} \label{}
\frac{d\sigma_{nl \rightarrow n'l'}}{d\Omega}
=\frac{1}{2l+1}\frac{k_f}{k_i}\sum_{m m'}|f(nlm\rightarrow
n'l'm'|\mathbf{k}_i, \mathbf{k}_f \rangle|^2,
\end{equation}
 partial cross sections
\begin{equation}
\sigma^J_{n l \rightarrow n'l'}( E) =
\frac{\pi}{k_{i}^2}\frac{2J+1}{2l+1} \sum_{L
L'}|T^J(nlL\rightarrow n'l'L')|^2, \label{}
\end{equation}
and the total cross sections for the  $n l\rightarrow n'l'$
transition are obtained by summing the corresponding partial
cross sections over the total angular momentum $J$:
\begin{equation}
\sigma_{nl \to n'l'}(E) = \sum_{J}\sigma^J_{n l \rightarrow
n'l'}( E).
\end{equation}

Finally, the averaged over the initial orbital angular
momentum $l$ cross sections for the $n \rightarrow
n'$ transitions  are then defined by summing over $l'$ and
$l$ ($l = l'$ for the elastic scattering and $l \neq l'$ for the
Stark transitions)  with the statistical weight
$(2l+1)/n^2$ in the case of the degenerated exotic atom
states and with the weight $(2l+1)/(n^2-1)$ in the case
when the energy shift of the $ns$ state is taken into
account:
\begin{equation}
\sigma_{n\to n'}(E) =
\frac{\pi}{k_{i}^2}\frac{1}{n^2}\sum_{l,\,l\,' \, L L'
J}(2J+1) |T^J(nlL\rightarrow n'l'L')|^2, \label{}
\end{equation}
and
\begin{equation}
\sigma_{n\to n'}^{l>0}(E) =
\frac{\pi}{k_{i}^2}\frac{1}{n^2-1}\sum_{l>0,\,l\,' \, L L'
J}(2J+1) |T^J(nlL\rightarrow n'l'L')|^2,
\label{}
\end{equation}
respectively.

\section{Results}

The close-coupling method described in the previous Section
has been used to obtain the total cross sections for the
collisions of the $\bar{p} p$ atoms in excited states with
hydrogen atoms. The present calculations had at least two goals:
first, to apply the fully quantum-mechanical approach for
the study of the processes (1) - (3) and, second, to clear
the effect of the energy shifts of the $ns$ states of the
antiprotonic hydrogen atom on the cross sections of these processes.

The coupled differential equations (16) are solved numerically
by the Numerov method  with the standing-wave boundary
conditions involving the real $K$-matrix. The corresponding
$T$-matrix are obtained from  $K$-matrix using the matrix
equation $ T=2iK(I- iK)^{-1}$. In the calculations both the exact 
interaction matrix and all the open channels with $n'\le n$ have
been taken into account. The closed channels are not
considered in the present study.
The close-coupling calculations have been carried out for
the relative collision energies $E_{\rm cm}$ from 0.05 up
to 50 eV and for the excited states with $n=3\div14$. At
all energies the convergence of the partial wave expansion
was achieved and all the cross sections were calculated with the
accuracy better than 0.1\%.

\begin{figure}[h]
\includegraphics[width=0.7\textwidth,keepaspectratio]{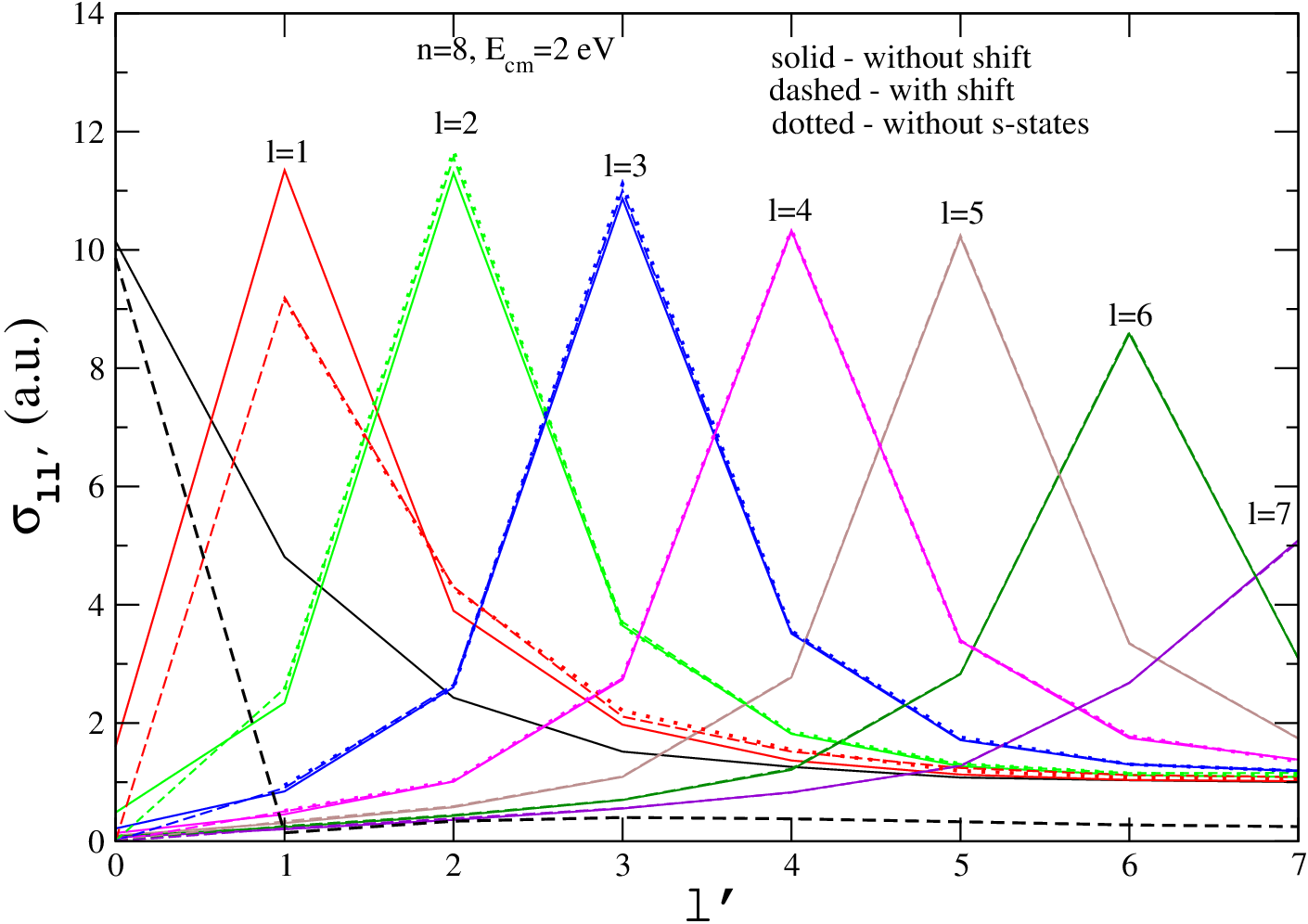}
     \caption{The total cross sections $\sigma_{nl\to nl'}$  for the 
     collisions of the $p\bar{p}$ atom in the $n=8$ state with 
     hydrogen atom at $E_{\rm cm}=1.4$~eV. The dashed and dotted 
     lines connect the points corresponding to the calculations both 
     with and without taking into account the $ns$-state energy shifts, 
     respectively. The dotted lines denote the results obtained
     without including the $ns$-states into the basis set.} 
      \label{fig1}
     \end{figure}

The results of the calculations are presented in Figures 1-4. 
In Fig.~1 we introduced the calculated total cross sections of the $nl \to nl'$ 
transitions for $n=8$ at the kinetic energy $E_{\rm cm}=1.4$~eV both with 
and without the $ns$ state energy shifts.
The following 
measured  world-average value~\cite{17} for the
spin-averaged shift 
$$\epsilon_{1s} = (721 \pm 14)  {\rm eV }$$ 
was used in the present calculations
and the energy shifts of the $ns$
states are defined by $\epsilon_{1s} /n^3$.

\begin{figure}[h!]
    \includegraphics[width=0.7\textwidth,keepaspectratio]{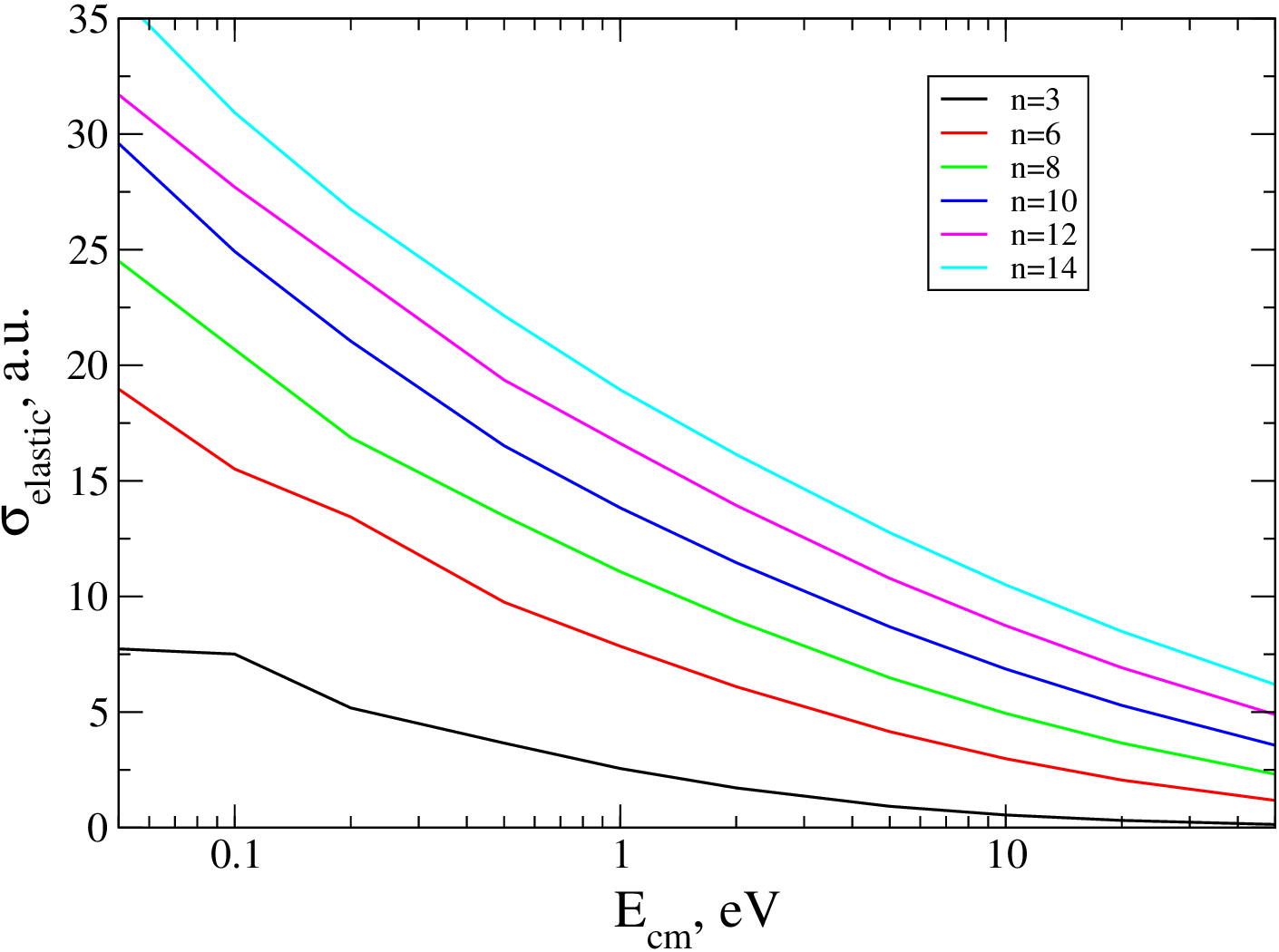}
     \caption{The $l$-averaged cross sections of the elastic scattering for 
     the collisions of the  $p\bar{p}$ atom with the hydrogenic atom for the 
     different values of the principal quantum number $n$}
 \label{fig2}
     \end{figure}

Contrary to the $(\pi^- p)$-atom, the energy shifts of the $ns$ states in 
$(\bar{p}p)$-atom are repulsive, hence,  
the $nl \to ns$ transitions are closed below the corresponding threshold and,
besides, according to the present study (e.g., see Fig. 1) the Stark transitions both
from the $ns$-states and to the  $ns$-state at the same collisional energy are strongly 
suppressed at the kinetic energies above the threshold. 
The similar effect is also observed for the
elastic $np \to np$ transitions.  The other transitions are practically
unchangeable at fixed energy. The analog of this effect can be modeled by 
excluding $ns$-states from the basis states. At energy above a few
thresholds the  effect is much weaker.  Therefore, the strong interaction effect
in the antiprotonic hydrogen (the similar effect must be also observed in the 
kaonic hydrogen atom) results  in the essential and important difference of the Stark transition
role in the absorption from the $ns$ states in contrast to pionic hydrogen
atoms. The influence of the strong interaction shift enhances for the lower states
and becomes less pronounced for the highly excited states of the antiprotonic atom. 

 \begin{figure}[h!]
 \includegraphics[width=0.7\textwidth,keepaspectratio]{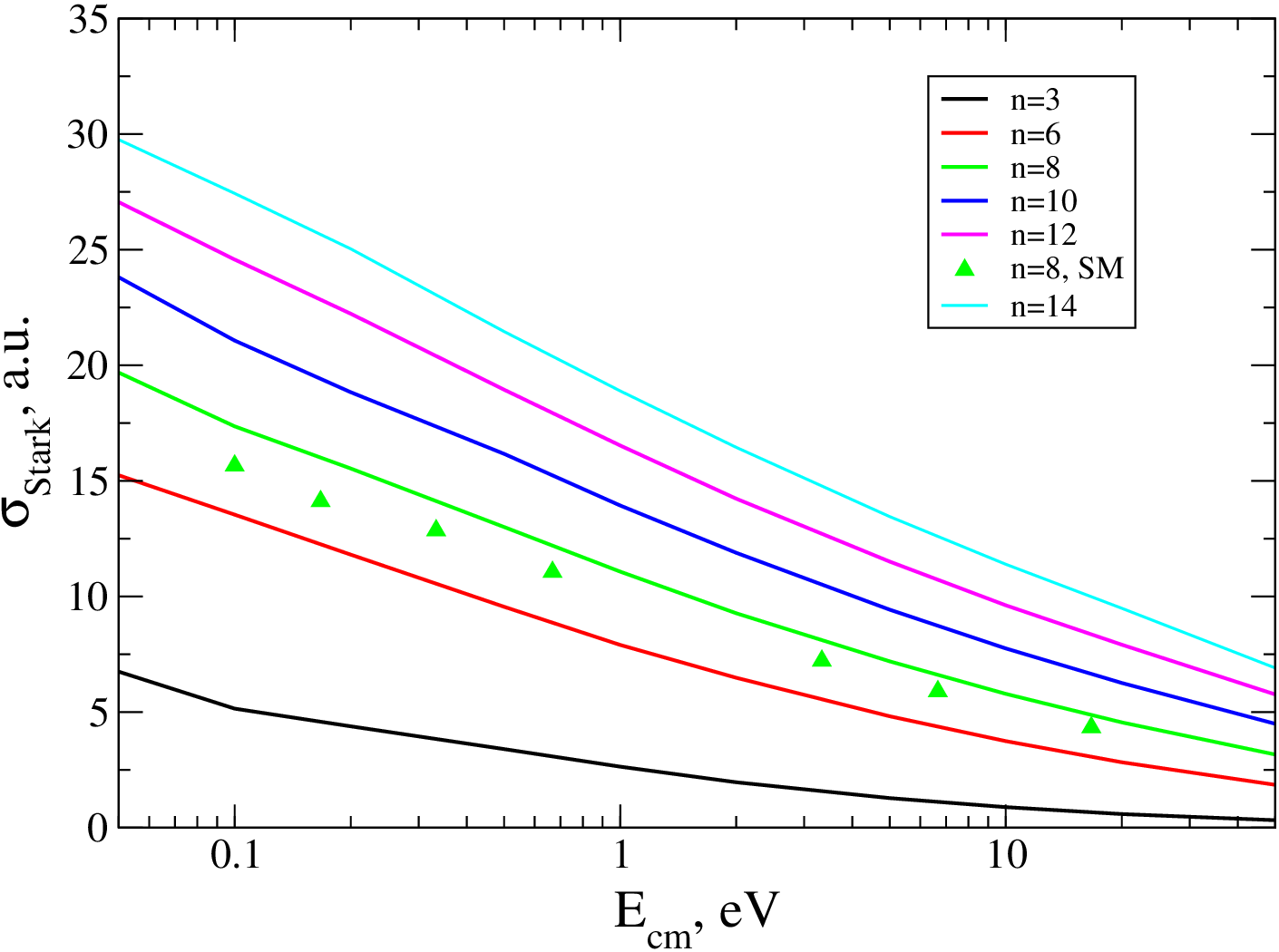}
  \caption{The $l$-averaged cross sections of the Stark transitions for 
  the collisions of the $p\bar{p}$ atom with the hydrogenic one. The results of 
  the calculations~\cite{10} in the semiclassical model are shown 
  for $n=8$ with triangles}
 \label{fig3}
     \end{figure}

 In Figures 2 and 3 the energy dependence of the $l$-averaged total elastic 
 and Stark cross sections are shown for the different principle
 quantum number values  from $n=3$ up to $n=12$. 
  Since the relative contribution of the $ns$ state in the $l$-averaged cross
 sections are small, the calculations both with and without the energy shift are
 practically indistinguishable at the energies above the corresponding threshold.
 So, in Figs. 2 and 3 the small energy region (below the corresponding
 threshold) corresponds to the calculations
 without the energy shift taken into account. In Fig. 3 the $l$-averaged cross
 section for $n$=8 are compared with the calculations in the framework of the
 semiclassical model~\cite{10}. As a whole a fair agreement is observed, 
 but the semiclassical model results in a different energy dependence
 especially at low energy collision and gives smaller cross sections than
 those obtained in the present approach. 

  \begin{figure}[h]
  \includegraphics[width=0.7\textwidth,keepaspectratio]{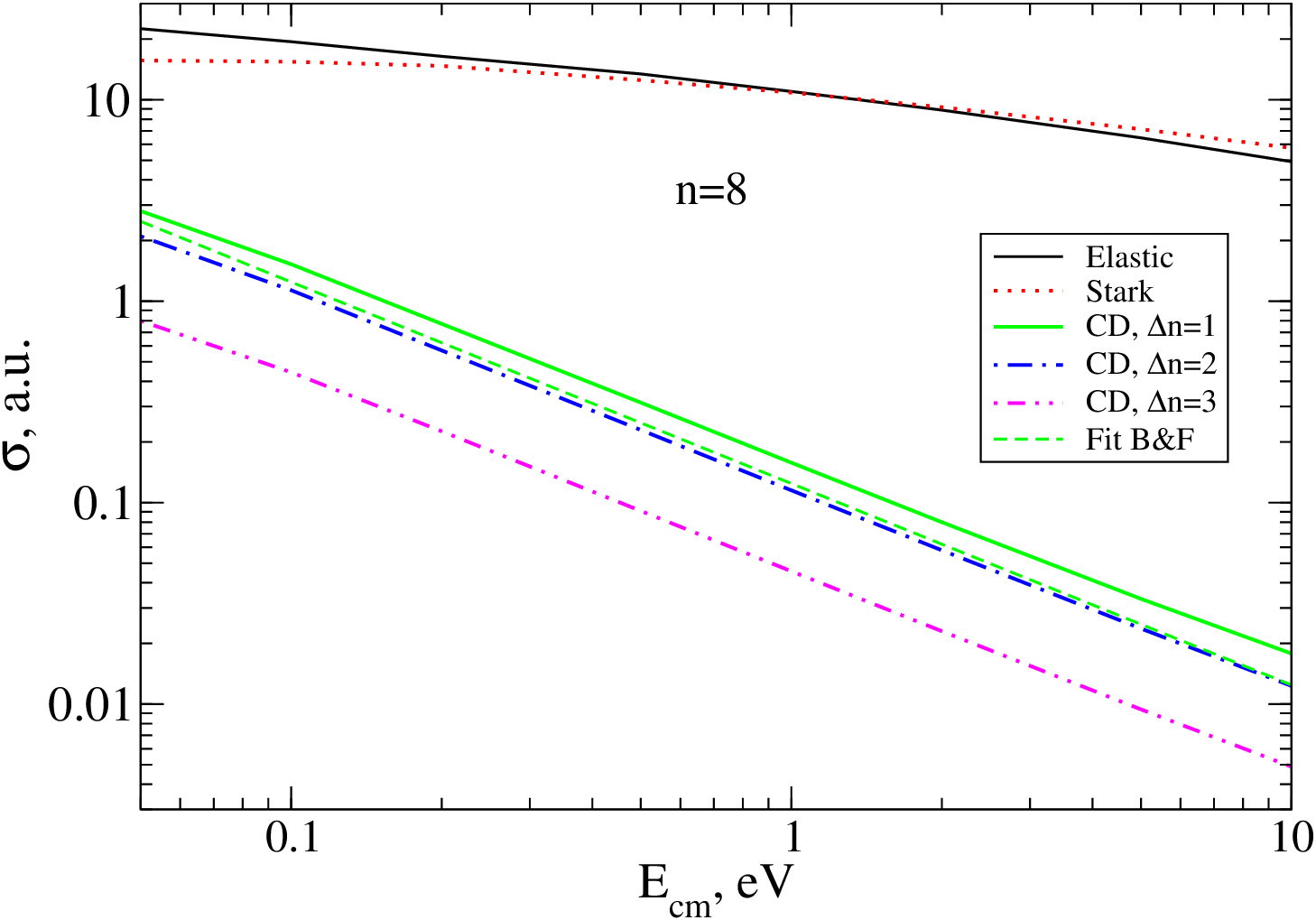}
  \caption{The $l$-averaged cross sections of Coulomb deexcitation with 
  $\Delta n=1,2,3$ for the collisions of the $p\bar{p}$ atom 
   ($n=8$) with the hydrogenic atom. The dashed line shows the fit 
   used in cascade calculations~\cite{13} for the transitions with 
   $\Delta n =1$and based on the mass-scaling of the results~\cite{11} for the
   muonic atom. The present calculations of the $l$-averaged elastic and 
   Stark cross sections are shown for comparison}
 \label{fig4}
     \end{figure} 
 
The dependence of the Coulomb deexcitation cross sections on energy obtained in
the present study is illustrated in Fig. 4 for $n=8$ and the different 
values of $\Delta n$ =1, 2 and 3. The special features of
these cross sections are the following: the similar energy dependence but sharper than that
of the elastic scattering and Stark transitions (see also in Fig. 4); 
the contribution of the transitions with  $\Delta n > 1$ is comparable with the
one for $\Delta n$ =1 and approximately equal about 50\%. The effect of the 
$ns$ state energy shifts in the $l$-averaged Coulomb deexcitation cross
sections are small for the same reason as it was discussed above (due to small
statistical weight of the ns-state). 
In Fig. 4 we also compare  our results with those obtained in the semiclassical model
(we use the parameterization  suggested in [13] which gives a fair description
of the Coulomb cross sections from [11])  for the $\Delta n$ =1 transition. 
The satisfactory agreement is observed, but  this agreement is
quite occasional and takes no place for other $n$ values. 
The distribution over the final states $n'$ is completely
different from the semiclassical results [11] as it was
illustrated in Fig.~4. The present calculations
predict that $\Delta n>1$ transitions give a substantial contribution to the
Coulomb deexcitation of the highly excited antiprotonic hydrogen atom that is in agreement
with our previous results for the muonic and pionic hydrogen~[14,15] atoms.

\section{Conclusion}
The unified treatment of the elastic scattering, Stark
transitions and Coulomb deexcitation is presented in {\em
ab initio} quantum-mechanical approach and for the first time 
the cross sections of these processes have been calculated for the 
highly excited antiprotonic hydrogen atom. 
The influence of the energy shifts of the $ns$ states on 
these processes has been studied.  
We have found that strong interaction shifts 
in the antiprotonic hydrogen atom lead to
substantial suppression of both the $ns \to nl'\neq 0$ and
$nl'\neq 0 \to ns$ transitions. At the same time the cross sections 
of the elastic scattering and Stark transitions for the the states with 
$l>2$ are practically unchangeable. 
The present study is the first step to achieving a reliable theoretical
input for realistic kinetics calculations.

We are grateful to Prof. L.Ponomarev for the stimulating
interest and  support of this investigation.

\end{document}